\documentstyle [12pt]{article}
\def\g5{\gamma^5}

\def\d4k{{d^4k\over (2\pi)^4}}

\newcommand{\beq}{\begin{eqnarray}}
\newcommand{\eeq}{\end{eqnarray}}
\newcommand{\beqno}{\begin{eqnarray*}}
\newcommand{\eeqno}{\end{eqnarray*}}
\def\lsim{\mathrel{\rlap{\lower4pt\hbox{\hskip1pt$\sim$}}
    \raise1pt\hbox{$<$}}}         
\def\gsim{\mathrel{\rlap{\lower4pt\hbox{\hskip1pt$\sim$}}
    \raise1pt\hbox{$>$}}}         

\begin{document}
%
\title{Vector, Axial, Tensor and Pseudoscalar Vacuum Susceptibilities}

\author{Leonard S. Kisslinger\\
        Department of Physics,\\
       Carnegie Mellon University, Pittsburgh, PA 15213}
\maketitle
\indent
\begin{abstract}
Using a recently developed three-point formalism within the method of QCD 
Sum Rules we determine the vacuum susceptibilities needed in the two-point 
formalism for the coupling of axial, vector, tensor and pseudoscalar currents
to hadrons.  All susceptibilities are determined by the space-time scale of
condensates, which is estimated from data for deep inelastic scattering on
nucleons.
\end{abstract}

\vspace{0.5 in}

\noindent
PACS Indices : 12.38.Lg, 13.30.Eg,13.40.Fn
\newpage
\section{Introduction}
\hspace{.5cm}

Since the operator product expansion (O.P.E.) fails for low momentum transfer,
the coupling of hadrons to currents cannot be treated by the usual
method of QCD sum rules extended to three-point functions. Recently, a 
three-point formalism, the standard starting point for such couplings in field
theory, was developed\cite{jk1} for a QCD sum rule approach to hadronic 
couplings with particular emphasis on vacuum susceptibilities.  It was shown 
that the vacuum susceptibilities that occur in the external field method
with two-point functions are given in the three-point approach by four-quark
correlations which can be estimated from the scale of the nonlocality of
quark condensates, with the main limitation of accuracy given by the use of
vacuum saturation. In Ref.\cite{jk1} the pion susceptibility was treated.
There has been a great deal of other work using three-point functions for
hadronic couplings [see Ref.\cite{jk1} for discussion and references].
In the present note the method is used to determine the susceptibilities for 
all currents that have been treated previously by QCD Sum Rules with the 
external field method.

In the early work on electromagetic coupling to nucleons\cite{by} it was
observed that the O.P.E. cannot be used consistently at low momentum transfer
and bilocal corrections were introduced.  This problem was solved using a
two-point function with an external field method\cite{is}, with
phenomenological vacuum susceptibilities introduced as parameters for
nonperturbative propagation in the external field.  Although the nucleon
magnatic dipole moments were determined in Ref.\cite{is} by sum rules which
avoided the use of the vector susceptibility, additional sum rules were used
to estimate the value of this parameter. There have been a number of
applications of this two-point method for the calculation of the axial
coupling constant (g$_A$) \cite{bk,cpw,hhk} and for the isoscalar axial
coupling constant (g$^S_A$)\cite{hhk}.  From the appropriate sum rules the
axial vector susceptibility has been estimated.  There as also been an
estimate of the vacuum tensor susceptibility\cite{hj} in work on the tensor
charge of the nucleon, however, it has been pointed out\cite{bo} that the
treatment of the nucleon tensor charge is subtle and that different
theoretical treatments can give very different results.

In a study of the parity-violating pion-nucleon interaction\cite{hhk1}
the pion-nucleon coupling was shown to depend critically on the pion-induced
susceptibility, which was estimated. Recently,
the axial-nucleon coupling, which involves the closely-related
pseudoscalar coupling constant was also treated in the external field QCD sum
rule method\cite {hhk2}. Although in principle the pion susceptibility can
be estimated using PCAC, in practice there are inconsistencies with the
values needed phenomenologically\cite{hhk1,jk1}

In the present note we show that the method developed in Ref.\cite{jk1}
can be successfuly used for predicting the induced vacuum suscetibilities 
for all of these currents. With the factorization approximation used in
our work there is a single scale provided by the space-time size of the
nonlocal condensates that determines all of the susceptiblities.

\section{Vacuum Susceptibilities}
\hspace{.5cm}

In a field theory the coupling of a current J$_\Gamma$(y) to a hadron
is studied by the three-point function:
\beq
  V^\Gamma(p,q) & = & \int d^4x \int d^4y e^{ix\cdot p} e^{-iy\cdot q} 
 V^\Gamma(x,y) \nonumber\\
 V^\Gamma(x,y) & = & <0|T[\eta(x) J^\Gamma(y)\bar \eta(0)]|0> 
\label{eq-vg2}
\eeq
where the quantity $\eta(x)$ is a composite field operator representing 
the hadron, with quark fields as constituents.  By evaluating V$^\Gamma$
at large p$^\mu$ one can make an operator product expansion with respect to
the (0,x) variables, and thereby introduce the nonperturbative QCD effects
through quark condensates.  However, since one wishes to evaluate
V$^\Gamma$ for small q, the O.P.E. for the (0,y) or (x,y) variable cannot
be justified and the QCD sum rule method for treating nonperturbative QCD
does not work.

  To avoid this difficulty a two-point formulation of the QCD Sum Rule in an
external electromagnetic field was introduced\cite{is}.  The form of the
correlator is
\beq
\Pi^\Gamma(p) & = 
  & i\int d^4 e^{ix\cdot p}<0|T[\eta(x)\bar\eta(0)]|0>_{J^\Gamma},
\label{eq-gpi}
\eeq
in which the presence of the external current J$^\Gamma$ is indicated by
the symbol in  Eq. (\ref{eq-gpi}).  By going to large p$^\mu$ the microscopic 
evaluation of $\Pi^\Gamma(p)$ can be done using the O.P.E of the quark
propagator in the presence of the the $J^\Gamma$ current 
\beq
 S^\Gamma_q(x) & = & <0|T[q(x)\bar{q}(0)]|0>_\Gamma, 
\nonumber\\
 	& = & S_q^{\Gamma,PT}(x) + S_q^{\Gamma,NP}(x),
\label{eq-gprop}
\eeq
where $S_q^{\Gamma,PT}(x)$ is the quark propagator coupled perturbatively to
the current and $S_q^{\Gamma,NP}(x)$ is the nonperturbative quark propagator
in the presence of the external current, $J^\Gamma$,
\beq
  S_q^{\Gamma,NP}(x)  & = & <0|:\bar q(x) q(0):|0>_{J^\Gamma}
\label{eq-sgnp}
\eeq
The operator product expansion for $S_q^{\Gamma,NP}(x)$ is justified as in
the ordinary two-point function.
The nonperturbative vacuum matrix elements for the quark propagator in
the external J$^\Gamma$ current, given in Eq.(\ref{eq-sgnp}), are expressed 
in terms of condensates.  For example, the lowest-dimensional term is
\beq
<0|:\bar{q} J^\Gamma q:|0>_\Gamma & = & - \Gamma \chi^\Gamma 
 <0|:\bar{q}q:|0>/12
\label{eq-chi1}
\eeq
in terms of the vacuum susceptibility $\chi^\Gamma$.
These susceptibilities must be determined in order to predict the
coupling constants using QCD sum rules.

In Ref.\cite{jk1} it was shown that $S_q^{cc'\Gamma,NP}(x)$ of
Eq.(\ref{eq-chi1}) with color indices c,c'is given in the three-point method
by the four-quark condensate
\beq
 S_q^{cc'\Gamma,NP}(x) & = & -i\int d^4y e^{-iy\cdot q} 
 <0|:q^c(x)\bar q^e(y) \Gamma q^e(y)\bar q^{c'}(0):|0>,
\label{eq-3p2p}
\eeq

If we assume vacuum saturation for intermediate states $S_q^{cc'\Gamma,NP}(x)$
is given in terms of nonlocal condensates, which have been used for the pion
wave function\cite{mr}, the pion form factor\cite{br} and deep inelastic
scattering from nucleon targets\cite{jk}. Taking the q$^\mu$ = 0 limit we
obtain 
\beq
S_{q}^{c c'\Gamma,NP}(x) & \simeq & \Gamma G(x) 
 (<0|:\bar q(0) q(0):|>/12)^2,
\nonumber \\
 G(x) & = & (-i)\int d^4y {\rm g}(y^2) {\rm g}((x-y)^2),
\label{eq-4qfac2}
\eeq
where the function g(y$^2$) gives the space-time structure of the nonlocal
condensates:
\beq
 <0|:\bar q(0) q(y):|0> & \equiv & {\rm g}(y^2) <0|:\bar q(0) q(0):|>,
\label{eq-nlc}
\eeq
In Ref.\cite{jk} it was shown that g(y$^2$) can be determined from the
experimental sea-quark distribution as measured in deep inelastic 
scattering on the nucleon.  Using the form
\beq
 {\rm g}(y^2) & = & \frac{1}{(1+\kappa^2y^2/8)^2} 
\label{eq-f}
\eeq
 we obtain
\beq
 S_{q}^{c c'\Gamma,NP}(x) & \simeq & \Gamma \frac{a^2}{6^3 \pi^2 \kappa^4}
\label{eq-3pchi}
\eeq
Finally, from a reanalysis of the QCD sum rule three-point treatment\cite{jk}
with the form given in Eq.(\ref{eq-f}) it is shown in Ref.\cite{jk1} 
$\kappa^2 \simeq$ (0.15-0.2) GeV$^2$.

\section{QCD Sum Rule Three-Pont Method for Known Susceptibilities}
\hspace{.5cm}

   In this section we apply the three-point method described in Sec. 2 for
the susceptibilities that have been estimated for axial vector, vector and
pionic currents and briefly discuss the tensor case.

\subsection{Axial Vector Susceptibility}
\hspace{.5cm}

   The axial and pseudoscalar coupling constants, g$_A$ and g$_P$, are 
defined by the nucleon matrix element of the axial current,
\beq
 <N(p')| J^5_\mu |N(p)> & = & \bar v(p')(g_A \gamma_\mu \gamma_5 +
  g_P q_\mu \gamma_5)v(p),
\label{eq-gp}
\eeq
In this subsection g$_A$ is discussed, while g$_P$ is discussed below.
There are two susceptibilities that appear in the conventional two-point
treatment of an external axial current, Z$_\mu$: $ \chi_A$ and $\kappa_A$ 
\beq
 <0|\bar q \gamma_\mu \gamma_5 q|0>_{Z}  & = & \chi_A
 <\bar q q> Z_\mu \\
\nonumber 
 <0|\bar q g_c \tilde{G}_{\mu\nu} \gamma^{\nu}  q|0>_{Z}  
 & = & \kappa_A <\bar q q> Z_\mu,
\label{eq-axsus}
\eeq
where g$_c$ is the strong coulpling constant and G$_{\mu\nu}$ represents
the gluon field in the fixed-point gauge and $\tilde{G}_{\mu\nu} =
\epsilon_{\mu\nu\alpha\beta}G^{\alpha\beta}$/2.  
The susceptibility $\chi_A$ 
involves the type of four quark matrix elements treated in the present work,
while the mixed susceptibility $\kappa_A$ involves some correlations not
considered here. In the treatment of the isovector axial coupling constant 
(g$_A$)\cite{bk,cpw,hhk} and the isoscalar axial coupling constant 
(g$^S_A$)\cite{hhk} the sum rule for the nucleon were used to   
reduced the dependence on these susceptibilities. We use the formulation
of Ref.\cite{hhk} here.  

The two sum rules in the two-point method are
\beq
 \beta_N^2(\frac{g_A}{M_B^2} + A) e^{-m^2/M_B^2} & = &
 \frac{M_B^4 E_2}{8 L^{4/9}} +\frac{< g_c^2 G^2 > E_0}{32 L^{4/9}}
+\frac{5 a^2 L^{4/9}}{18 M_B^2} -\frac{\kappa_A a E_{0}}{18 L^{68/81}} 
\nonumber\\
   & & +\frac{\chi_A a < g_c^2 G^2 > E_0}{288 M_B^2 L^{4/9}}  
\label{eq-ss1}
\eeq
and
\beq
 \beta_N^2 g_A(1-\frac{2m^2}{M_B^2}) e^{-m^2/M_B^2} & = &
 \frac{M_B^6 E_2}{8 L^{4/9}} +\frac{M_B^2 < g_c^2 G^2 > E_0}{32 L^{4/9}}
+\frac{a^2 L^{4/9}}{18 M_B^2} -\frac{M_B^2 \kappa_A a E_{0}}{2 L^{68/81}}
\nonumber\\
   & & +\frac{\chi_A a < g_c^2 G^2 > E_0}{288 M_B^2 L^{4/9}} 
\label{eq-ss2},
\eeq
with M$_B$ the Borel mass, $< g_c^2 G^2 > \simeq$ 0.47 GeV$^4$, 
a = -(2$\pi$)$^2$$<\bar q q>$ $\simeq$ 0.55 GeV$^2$, L = .62 ln(10 M$_B$),
and $<0|\eta(x)|N>$ = ($\beta_N $ /$\pi^2$) v(x), where v(x) is a Dirac
spinor. 
The functions E$_m$, defined in Ref.\cite{hhk}, and the parameter A
help treat the continuum.

In the evaluation of these sum rules in Ref.\cite{hhk} it was shown 
that g$_A$ -1 is mainly given by the quark
condensate term proportional to a$^2$ with a value of 0.25, in agreement
with experiment and the Goldberger-Trieman relation, and giving the 
expected result of g$_A \simeq$ 1.0 when the quark condensates vanish with
chiral symmetry restoration.  The susceptibility terms
were unimportant in that evaluation.  Here we eliminate the susceptibility
$\kappa_A$ from Eqs.(13,14) to obtain an expression for the
susceptibility of interest, $\chi_A$.  We find that
\beq
  \chi_A(two-point) a & \simeq & -1.24 GeV^2 
\label{eq-chiA2}
\eeq
This value is in agreement within expected errors of the value predicted from
the three-point method (with G(0) = $2^5\pi^2/3\kappa^4)$:
\beq
  \chi_A(three-point) a & \simeq & -G(0)\frac{a^2}{12(2\pi)^2}\\
\nonumber
  & \simeq & -(1.7-3.0) GeV^2. 
\label{eq-chiA3}
\eeq
Since the susceptibility gives a rather minor contribution for the isovector
axail coupling constant, the value given in Eq.(\ref{eq-chiA2}) is not very
accurate, but shows that the prediction of the three-point method with no
free parameters is consistent with the value extracted by the two-point
method.  The study of g$_A^S$, the isoscalar axial coupling
constant\cite{hhk} adds no new information.  The prediction of the present
work would be the same for $\chi_{A^S}$ and in Ref.\cite{hhk} the same values
were used for the isoscalar susceptibilities as for the isovector case.

\subsection{Vector Susceptibility}
\hspace{.5cm}

Susceptibilities associated with the electromagnetic field were treated
by Ioffe and Smilga\cite{is} in their pioneering work on the use of the
two-point QCD sum rule method for calculation of nucleon magnetic moments.  
Using the fixed-point gauge and representing the electromagnetic field by the
F$_{\mu\nu}$ electromagnetic tensor, one finds three electromagnetic 
susceptibilities
\beq
 <0|\bar q \sigma_{\mu\nu} q|0>_F  & = & e_q\chi_V <\bar q q> F_{\mu\nu},\\
\nonumber 
 <0|\bar q g_c G_{\mu\nu} q|0>_F  & = & e_q\kappa_V <\bar q q> F_{\mu\nu},\\
\nonumber
 <0|\bar q g_c \tilde {G}_{\mu\nu} q|0>_F  & = & e_q\xi_V <\bar q q> 
 F_{\mu\nu},
\label{eq-vsus}
\eeq
As in the axial case, only the susceptibility $\chi_V$ is treated here.
There are three sum rules, two of which were used in Ref.\cite{is}.  Although
it is possible to make estimates of the nucleon moments by eliminating the
susceptibilities, they give quite important effects for the nucleon moments.  
A detailed analysis resulted in the estimate
\beq
  \chi_V(two-point) a m & \simeq & -(3.5-6.1) GeV^2 ,
\label{eq-chiV2}
\eeq
while the present three-point sum rule method gives
\beq
  \chi_V(three-point) a m & \simeq & -(3.4-6.0) GeV^2.
\label{eq-chiv3}
\eeq
Once more the prediction of the present theory is in satisfactory agreement
with the value for the susceptibility extracted from the two-point external
field method and experiment.

\subsection{Pion and Pseudoscalar Susceptibilities}
\hspace{.5cm}

The susceptibility associated with an external pion field was treated in
Ref.\cite{jk1}.  We briefly review this here. The strong and parity-violating
pion-nucleon coupling depends crucially upon $\chi_\pi $, the pion
susceptibility\cite{hhk1}
\beq
 <0|\bar q \tau_3\gamma_5 q|0>_\pi  & = &  i g_\pi \chi_\pi <\bar q q>
\gamma_5
\label{eq-pisus}
\eeq
From the analysis of the strong pion-nucleon coupling constant, which is
known to be g$_{\pi}-N$ = 13.5, one finds in the two-point QCD sum rule
method
\beq
 \chi_\pi(two-point) a  & \simeq & 1.88 GeV^2.
\label{eq-chipi2}
\eeq
Using the three-point QCD sum rule method one finds\cite{jk1}
\beq
 \chi_\pi(three-point) a  & \simeq & \frac{2a^2}{9 \kappa^4}\\
\nonumber
   & \simeq & (1.7-3.0) GeV^2,
\label{eq-chipi3}
\eeq
in agreement with the value extracted by the external field method.

   The pseudoscalar coupling, g$_P$ is difficult to treat by the two-point
external field QCD sum rule method designed for zero momentum transfer, as
one can see from Eq. (\ref{eq-gp}). 
Recently we have carried out such a study and have shown the crucial
dependence on susceptibilities\cite{hhk2}. In order to derive sum rules for
g$_P$ one can introduce an external field of the form
\beq
  Z_\mu(x) & = & Z_{\mu\nu}x^\nu,
\label{eq-Z}
\eeq
which in turn leads to the pseudoscalar vacuum susceptibility, $\chi_P$,
defined in momentum space at low momentum transfer by
\beq
 <0|\bar q \gamma_5 q|0>_{Z_\mu}  & = &  i  \chi_P <\bar q q> Z_{\mu\nu}
q_\nu\gamma_5.
\label{eq-psus}
\eeq
The three-point formalism must be modified so that Eq.(\ref{eq-4qfac2})
becomes
\beq
S_{q}^{c c'P,NP}(x) & \simeq & \gamma_5 Z^{\mu\nu} G_\nu(x) 
 (<0|:\bar q(0) q(0):|>/12)^2,
\nonumber \\
 G_\nu(0) & \simeq & Lim_{q \rightarrow 0} q_\nu\int d^4y {\rm g}^2(y^2)
 e^{iq\cdot y}\nonumber \\
          & = & q_\nu 4 G(0)/\kappa^2.
\label{eq-sp}
\eeq
From Eq.(\ref{eq-sp}) one finds
\beq
 g M \chi_P(three-point) a & \simeq & (40-53) GeV,
\label{eq-chip3}
\eeq
corresponding to the space-time range of the nonlocality given by
$0.2> \kappa^2 > 0.15$. This range of values for $\chi_P$ is found to
predict\cite{hhk2} values for g$_P$ consistent with the Goldberger-
Trieman relation.
\subsection{Tensor Susceptibility}
\hspace{.5cm}

The tensor susceptibility is defined through Eq.(\ref{eq-4qfac2}) and
\beq
 <0|\bar q \sigma_{\mu\nu} q|0>_Z  & = & e_q\chi_T <\bar q q> Z_{\mu\nu},
\label{eq-chitensor}
\eeq
with $\chi_T$ a factor 6 larger than the definition in Ref.\cite{hj}.
We obtain
\beq
  \chi_T(three-point) a/6  & \simeq & -(0.57-1.0) GeV^2.
\label{eq-chit3}
\eeq 
Two estimates of $\chi_T$ made in Ref.\cite{hj} were about -0.15 GeV$^2$
and -0.29 GeV$^2$, much smaller that the values that we find; however,
the authors point out that the estimates are crude and that the
susceptibilities do not give large effects for the tensor charge of the
nucleon.

\section{Conclusions}
\hspace{.5cm}

The three-point QCD sum rule method of Ref.\cite{jk1} with the 
factorization approximation has been used to estimate the external-field
susceptibilities which appear in the external field two-point method in terms
of a parameter giving the space-time scale of the quark condensate. 
Satisfactory agreement with all know vacuum susceptibilities has been found.
There is one scale, the size of the space-time structure of the nonlocal
condensates, which determmines the susceptibilities within our approximations.
This new method, using nonlocal condensates, does not require an operator 
product expansion of the quark propagators and can be used to study coupling
constants and hadronic form factors as a function of momentum transfer.

The author would like to acknowledge many helpful discussion with Drs.
Mikkel Johnson, Ernest Henley, Pauchy Hwang, and theorists at the Beijing
Institute of High Energy Physics on various aspects of this work.
The work was supported in part by the National Science Foundation
grants PHY-9722143 and INT-9514190.

\end{document}